\documentclass[prl,preprint,superscriptaddress]{revtex4}
\usepackage{dcolumn}
\usepackage{mathtools}
\usepackage{amssymb}
\usepackage{bm}
\usepackage{pifont}
\usepackage{psfrag}
\usepackage{epstopdf}
\usepackage{wasysym}
\usepackage{amsmath}
\usepackage{tipa}
\usepackage[usenames]{color}
\begin{document}

\title{Emergent Stereoselective Interactions and Self-recognition in Polar Chiral Active Ellipsoids}
\author{Pragya Arora}
\email{pragyaarora26@gmail.com}
\affiliation{Chemistry and Physics of Materials Unit, Jawaharlal Nehru Centre for Advanced Scientific Research, Jakkur, Bangalore - 560064, INDIA}
\author{A K Sood}
\affiliation{Department of Physics, Indian Institute of Science, Bangalore- 560012, INDIA}
\affiliation{International Centre for Materials Science, Jawaharlal Nehru Centre for Advanced Scientific Research, Jakkur, Bangalore - 560064, INDIA}
\author{Rajesh Ganapathy}
\email{rajeshg@jncasr.ac.in}
\affiliation{International Centre for Materials Science, Jawaharlal Nehru Centre for Advanced Scientific Research, Jakkur, Bangalore - 560064, INDIA}
\affiliation{School of Advanced Materials (SAMat), Jawaharlal Nehru Centre for Advanced Scientific Research, Jakkur, Bangalore - 560064, INDIA}

\date{\today}

\begin{abstract}
\textbf{
In many active matter systems, particle trajectories have a well-defined handedness or chirality. Whether such chiral activity can introduce stereoselective interactions between particles is not known. Here we developed a strategy to tune the nature of chiral activity of 3D-printed granular ellipsoids without altering their shape or size. In vertically agitated monolayers of these particles, we observed two types of dimers form depending on the chirality of the pairing monomers. Heterochiral dimers moved collectively as a single achiral active unit, while homochiral ones formed a translationally immobile spinner. In active racemic mixtures, the former was more abundant than the latter indicating stereoselectivity. Through dimer lifetime measurements, we provide compelling evidence for chiral self-recognition in mixtures of particles with different chiral activities. We finally show that changing only the net chirality of a dense active liquid from a racemic mixture to an enantiopure liquid fundamentally alters its nature of collective relaxation. }
\end{abstract}
 
\maketitle
When chirality is present as a static feature of building blocks, interactions between and reactions involving them are often stereoselective \cite{morrow2017transmission}. In fact, for molecular building blocks, chirality also leads to higher-order effects such as, self-recognition, sorting and discrimination \cite{jedrzejewska2017making}. There are however many instances where chirality is manifest, sometimes entirely, in the dynamics. Examples in the realm of classical physics where particle trajectories have a chirality/handedness associated with them include - assemblies of spinning colloidal magnets \cite{kokot2017active,soni2019odd} and granules \cite{grzybowski2002dynamic}, spinning \cite{petroff2015fast} as well as circular \cite{jennings1901significance} and helical swimmers \cite{su2012high}, and synthetic chiral active particles \cite{tsai2005chiral,kummel2013circular,workamp2018symmetry,scholz2018rotating}. In one of these systems - magnetic disks set spinning by a rotating external field - where chirality was present in both particle shape and dynamics, stereoselective interactions between particles aided in their dynamic self-assembly \cite{grzybowski2002dynamic}. A fundamental question that has remained unanswered hitherto is whether chirality in dynamics alone can induce similar effects.  In an externally driven assembly of shape \textit{achiral} active rotors/spinners, because all particles have the same handedness in their dynamics, stereoselectivity cannot arise. In active matter \cite{marchetti2013hydrodynamics}, on the other hand, the direction of motion, also the handedness if activity is chiral \cite{lowen2016chirality}, is set by the particle and this results in rich dynamical behaviour \cite{tsai2005chiral,liebchen2017collective,scholz2018rotating,lei2019nonequilibrium}. In fact, even without attractive interparticle interactions, activity, even when achiral, can result in condensed phases \cite{palacci2013living,cates2015motility}.  These systems are therefore a fertile playground to probe chiral activity-mediated effects. 

Addressing this question in experiments, even in 2D, requires a system that provides some degree of control over chiral activity; in 2D the relevant parameters are the radius of the circular trajectory, $R$, and the angular velocity, $\omega$, of the particle \cite{lowen2016chirality}. Tuning either or both is difficult in living chiral active matter. While synthetic active spinners are easy to fabricate \cite{tsai2005chiral,workamp2018symmetry,scholz2018rotating}, being subject to only an active torque, $R=0$, and this constrains the parameter space of chiral activity. To realize the more generic circular trajectories, where both active torques and forces act, existing strategies for both wet and dry active matter rely on particle shapes that are chiral \cite{tsai2005chiral,kummel2013circular}. This makes it impossible to disentangle emergent effects associated with chiral activity from those arising due to the chiral particle shape.

We achieved tunable chiral active motion while keeping the particle shape and size fixed and achiral, by capitalizing on a hitherto unexploited  feature of a canonical 2D active matter system - assemblies of millimeter-sized grains rendered active through vertical agitation. Previous experiments found that vibrated granules of simple shapes, like disks or rods, but with an asymmetry in mass, $m$, and/or friction coefficient, $\mu$, were polar active along the direction set by the asymmetry \cite{kudrolli2008swarming,deseigne2010collective,kumar2014flocking}. While in these studies, when the asymmetry in both were present they were coincident, in our granules - 3D-printed plastic ellipsoids - over and above an asymmetry in $\mu$ and $m$ along the major-axis, we also had an asymmetry in $m$ along the minor-axis. We anticipated vertical agitation to generate both an active force and torque on these particles and converged on this specific design after systematically excluding out other designs based on the nature of active dynamics observed (see Fig. S4). 

\begin{figure}
\includegraphics[width=1\textwidth]{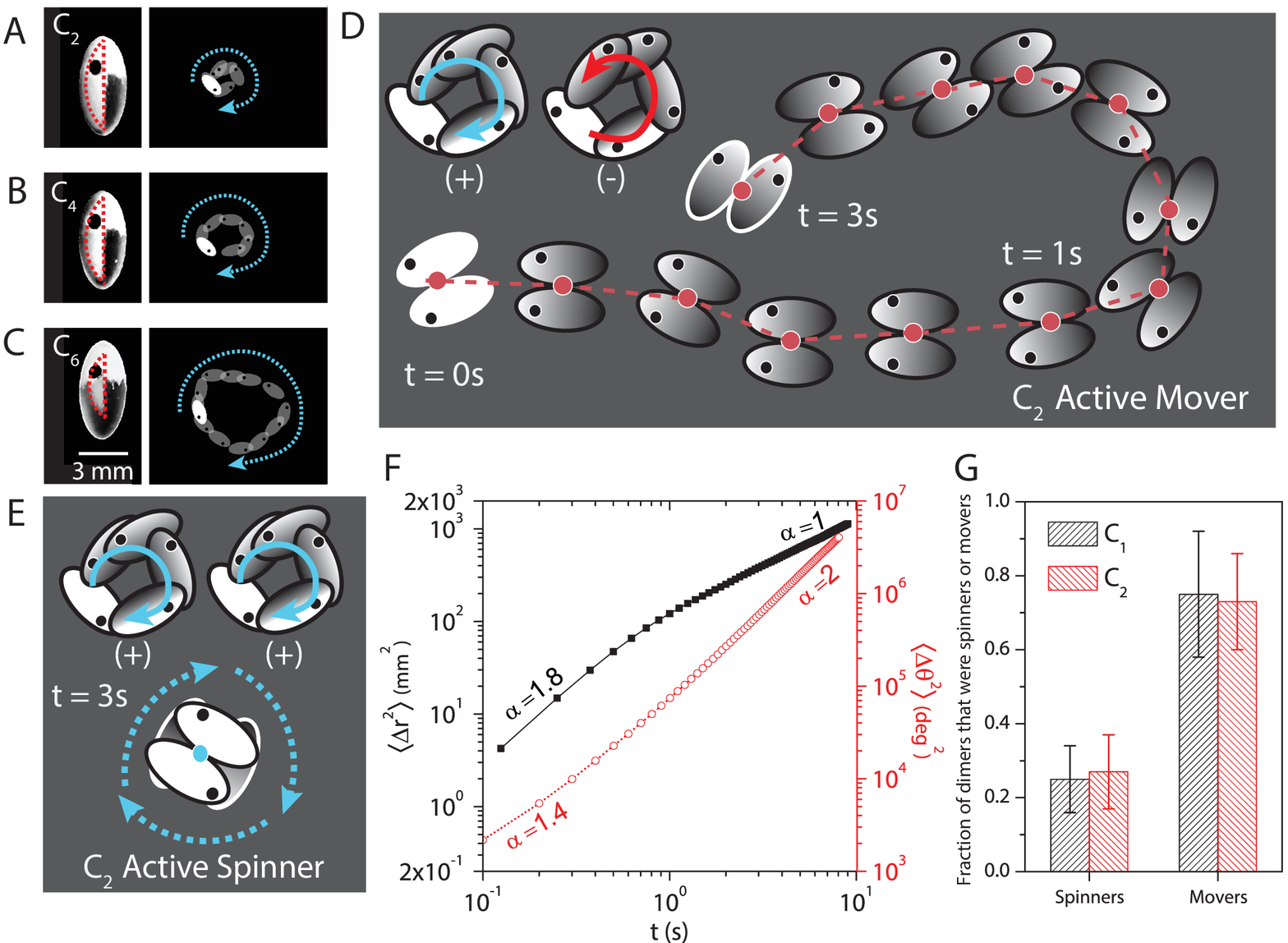}
	\caption{\textbf{Emergent stereoselective interactions in polar chiral active ellipsoids.} \textbf{(A-C)} Left panels: Snapshots of 3D printed chiral active ellipsoids for three different left-right mass asymmetries $\Delta m_{LR}$. The red dashed lines show the hollowed out portion of the particle. Right panels: Superimposed snapshots showing a nearly circular path traced by the ellipsoids under vertical agitation. The snapshot of the ellipsoid at $t=0$ s is shown in white. The blue arrow indicates the handedness of the orbit. \textbf{(D) and (E)} Superimposed snapshots of a representative active `mover' and `spinner', respectively. The mover is composed of  a dextro- (+) and levogyre (-) monomer (top left in \textbf{(D)}) while the spinner is made of two $(+)$ monomers (\textbf{(E)}). Note, the spinner has a net clockwise (+) motion (blue dashed arrow) same as that of its components and is localised in space. \textbf{(F)} Translational mean-squared displacement (MSD) of mover (black squares) and orientational MSD of spinner (red circles) versus lag time $t$. \textbf{(G)} Fraction of dimers that existed as movers and spinners in liquids of $\text{C}_{1}$ as well as $\text{C}_{2}$ ellipsoids at $\phi=1\%$. The error bars represent standard deviation of the mean and were obtained from multiple statistically independent realizations of an experiment.}
\label{Figure1} 
\end{figure}

Our Janus ellipsoids had a major-axis $\alpha = 6$mm and minor-axes $\beta = 3$ mm, $\gamma = 2.1$ mm. We exploited a feature  unique to the 3D-print process to introduce a fore-aft asymmetry in $\mu$ and this can be seen as a difference in the surface finish between the lower and upper half of the ellipsoid in Fig. \ref{Figure1}A-C. A left-right asymmetry in $m$, with respect to the particle major-axis, was achieved by making one portion of the ellipsoid hollow during print  (shown by dashed red line in Fig. \ref{Figure1}A-C, also see Fig.  S4). The printed particles were placed on a horizontally mounted flower-shaped arena and confined from above with a transparent glass plate which also enabled imaging of dynamics. The entire assembly was coupled to a permanent magnet shaker through a stiff air-bearing which ensured that the imposed sinusoidal driving from the shaker is channeled only into vertical oscillations of the plate\cite{harris2015generating}(see Materials and Methods). The drive frequency $f = 37$ Hz and amplitude $a  = 1$ mm were kept constant in all experiments, unless specified, and the non-dimensional acceleration $\Gamma = {4\pi^2f^2a\over g} = 5.5$, where $g$ is the acceleration due to gravity. We quantified the dynamics of individual ellipsoids by working at a very low area fraction $\phi\approx 0.1\%$ ($N \approx 8-10$ particles on the plate), where $\phi$ is the total projected area of all particles on the plate divided by the surface area of the plate. The energy gained by the particles through frequent collisions with the vibrating plate resulted in polar chiral active motion - noisy circular trajectories with a more or less well-defined $R$ and $\omega$  (Fig. \ref{Figure1}A right panels and Movie S1). Since no special care was taken while placing particles on the plate, the granular assemblies were racemic mixtures and had nearly equal numbers of particles with dextrogyre ($(+)$) and levogyre ($(-)$) trajectories. Further, as the gap between plates $\Delta = 2.4$mm is such that $\gamma<\Delta<\beta$, particles cannot flip once they are confined and the handedness of the orbits remained unchanged. 

Now, by changing only the volume of the hollowed-out portion of the ellipsoid during 3D-printing (red dashed lines in Fig. \ref{Figure1}B and C) we tuned the extent of left-right mass asymmetry, $\Delta m_{LR}$, and hence the active torque (see Fig. S5 and Fig. S6).  As anticipated, decreasing $\Delta m_{LR}$ resulted in a smaller active torque and thus a decrease in $\omega$, and an increase in $R$ (right panels in Fig. \ref{Figure1}A-C and Fig. S7). Ellipsoids with different $\Delta m_{LR}$ values are labeled $\text{C}_1 - \text{C}_6$, where a larger subscript indicates a larger $R$ (see Table \ref{Table1}).

Having gained the ability to tune chiral activity, we  investigated dynamics of the granular assemblies at higher $\phi$. Even at a $\phi$ as low as $1\%$, we observed dimer formation solely due to chiral activity. The more conventional motility-induced clustering of achiral active particles sets in at a much larger $\phi$, typically in excess of $20\%$ \cite{cates2015motility}. We observed qualitatively similar dynamics for $\text{C}_1$ and $\text{C}_2$ ellipsoids. In racemic mixtures of these particles, we observed two distinct types of active dimers which we classified as `movers' and `spinners' based on their dynamics. Remarkably, this difference in behaviour was entirely a consequence of their composition. A $(+)$ and a $(-)$ monomer with their polar axis pointing in nearly the same direction, upon collision, formed an adduct that was torque free and behaved like an achiral polar active particle - a `mover' (Movie S2). Figure \ref{Figure1}D shows the representative experimental trajectory of a mover made of $\text{C}_2$ monomers. As is typical of an achiral active particle, the translational mean-squared displacement of the centre-of-mass (COM) of the adduct was superdiffusive at small lag times $t$ and then crossed over to diffusive dynamics at larger $t$ (solid black symbols in Fig. \ref{Figure1}F) \cite{bechinger2016active}.  In stark contrast, monomers that were both $(+)$ or both $(-)$, but whose polar-axis pointed in opposite directions, upon collision, formed a `spinner' with the same handedness (see Fig. \ref{Figure1}E and Movie S3 for a spinner made of $\text{C}_2$ monomers). The orientational MSD of the spinners showed a crossover from super-diffusive to ballistic behaviour with $t$ (red circles in Fig. \ref{Figure1}F). Due to the almost complete cancellation of the active forces, the translational dynamics of the spinner was purely diffusive with a diffusion constant that was nearly two orders-of-magnitude smaller than that of the movers (Fig. S8).

Interestingly, such mover and spinner states are theoretically predicted for two hydrodynamically interacting active rotors \cite{fily2012cooperative}. However, hydrodynamic interactions being long-ranged, whether these pairings survive at higher $\phi$ is not clear, and it is also not known if in a racemic mixture one of these bound states is more abundant, which would then indicate active stereoselectivity \cite{jedrzejewska2017making}. To determine if in our system interactions were stereoselective, we calculated the fraction of dimers that  existed as movers and spinners. In racemic mixtures of  $\text{C}_1$ as well as $\text{C}_2$ ellipsoids, the mover fraction is nearly three times the spinner fraction - a clear sign of active stereoselectivity (Fig. \ref{Figure1}G). For larger $R$ ($\text{C}_3$ to $\text{C}_6$ ellipsoids), we exclusively observed only movers although its lifetime decreased with $R$ since the opposing active torques keeping the monomers together also decreased (Movie S4). The absence of spinners at large $R$ is also understandable. For a spinner to be stable, monomers should not slip past each other and this is possible only when $R$ is comparable to the radius of revolution of the monomer about the spinners' COM, which is nothing but $\beta\over 2$.  Indeed, $R<$ $\beta\over 2$ for $\text{C}_1$ and $\text{C}_2$ ellipsoids and $R>$ $\beta\over 2$ for $\text{C}_3-\text{C}_6$ ellipsoids (Table \ref{Table1}).

We now considered the possibility of this active stereoselective interaction leading to chiral self-recognition  \cite{jedrzejewska2017making}, i.e., in assemblies made of ellipsoids of very different chiral activities, there was a preference in the way particles paired-up. We focused on mixtures of $\text{C}_2$ and $\text{C}_4$ ellipsoids as their dynamics were representative of the limiting cases $R<$ $\beta\over 2$ and $R>$ $\beta\over 2$, respectively. Movie S5 shows the dynamics of a $1:1$ mixture of individually racemic $\text{C}_2$ and $\text{C}_4$ ellipsoids at $\phi = 10\%$. This experiment provided the first cue that there might indeed be some form of self-recognition between particles - we observed $\text{C}_2\text{C}_2$ spinners and never any $\text{C}_4\text{C}_4$ spinners, as expected, but  we also never saw any $\text{C}_2\text{C}_4$ spinners. We however observed $\text{C}_2\text{C}_2$, $\text{C}_4\text{C}_4$ and also $\text{C}_2\text{C}_4$ movers. Measuring their average lifetimes to determine if there was self-recognition was difficult at high $\phi$ due to their interactions with other particles and at low $\phi$ we could not obtain enough statistics.

\begin{figure}[tbp]
\includegraphics[width=1\textwidth]{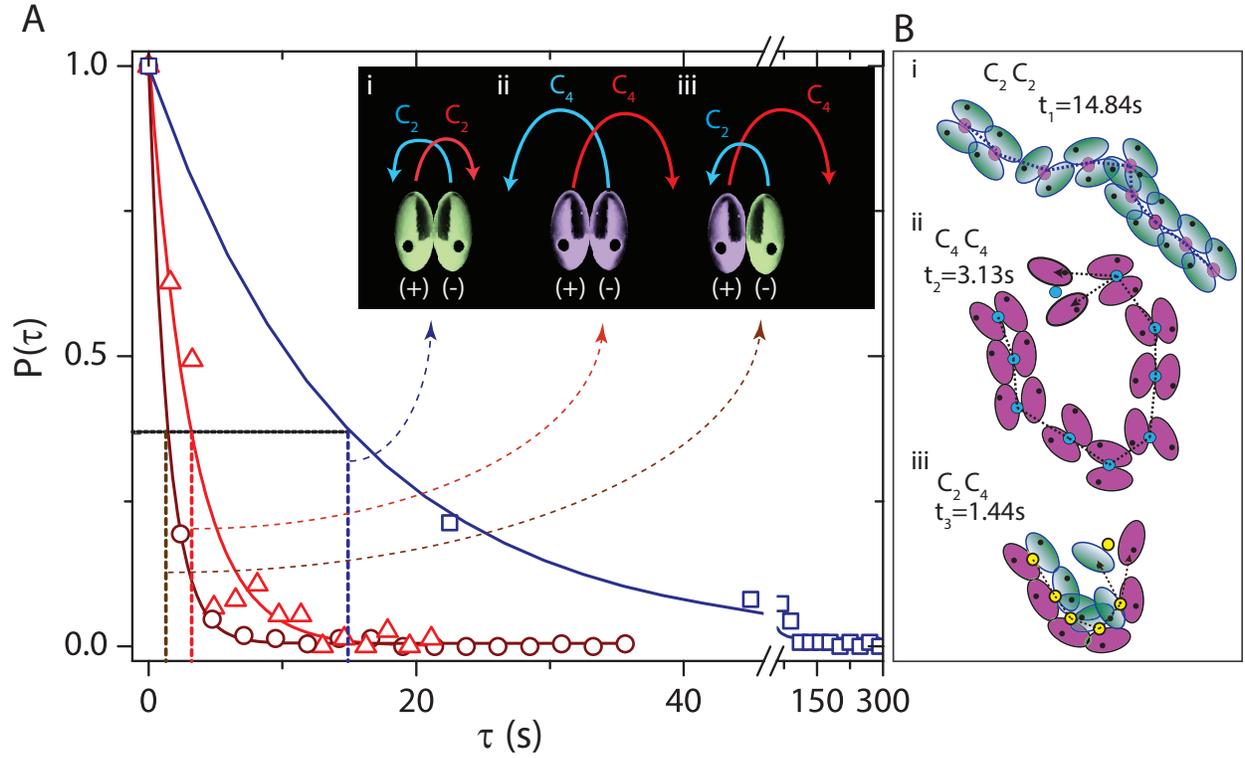}
\caption{\textbf{Self-recognition in polar chiral active ellipsoids.} \textbf{(A)}Probability distribution of mover lifetimes for the three possible configurations shown in the inset. \textbf{(i)}  $\text{C}_{2}\text{C}_{2}$ (hollow blue squares), \textbf{(ii)} $\text{C}_{4}\text{C}_{4}$ (red triangles) and \textbf{(iii)} $\text{C}_{2}\text{C}_{4}$(brown circles). Both the radius (represented by the arrows in the inset) and the angular velocities of  $\text{C}_{2}$ and $\text{C}_{4}$ are different. \textbf{(B)} Superimposed snapshots of representative active mover trajectory for each of the three configurations. The average lifetime $\tau_{mov}$ (shown by dashed vertical lines in (\textbf{A}) of $\text{C}_{2}\text{C}_{2}$ mover is 14.8 s ,  $\text{C}_{4}\text{C}_{4}$ mover is 3.1 s and $\text{C}_{2}\text{C}_{4}$ mover is 1.4 s.} 
\label{Figure2}
\end{figure}

To overcome this difficulty, before confining the particles and setting the plate in motion, we manually created one each of the three mover configurations - $\text{C}_2\text{C}_2$, $\text{C}_4\text{C}_4$ and $\text{C}_2\text{C}_4$ - by placing the appropriate $(+)$ and $(-)$ monomers side-by-side (inset to Fig. \ref{Figure2}A). This initial configuration helped significantly promote dimerization under vertical agitation (Movie S6). Trajectories representative of each of the three movers is shown in Fig. \ref{Figure2}B. By repeating this experiment a few hundred times, we measured the probability distribution of mover lifetimes, $P(\tau)$, for each of the three movers.  $P(\tau)$ was an exponential for all dimer types and we defined the average lifetime as $P(\tau = \tau_{mov}) = {1\over e}$. These measurements provided the first evidence for chiral self-recognition in active matter: movers made of like particles were longer-lived in comparison to those made of unlike ones (Fig. \ref{Figure2}A). In $\text{C}_2\text{C}_4$ movers, besides the radii, the angular velocities of the constituent monomers are also different (Table \ref{Table1}). The monomers thus slide past each other and the lifetime of these movers is smaller than even $\text{C}_4\text{C}_4$ ones.

\begin{figure}[tbp]
\centering
\includegraphics[width=0.85\textwidth]{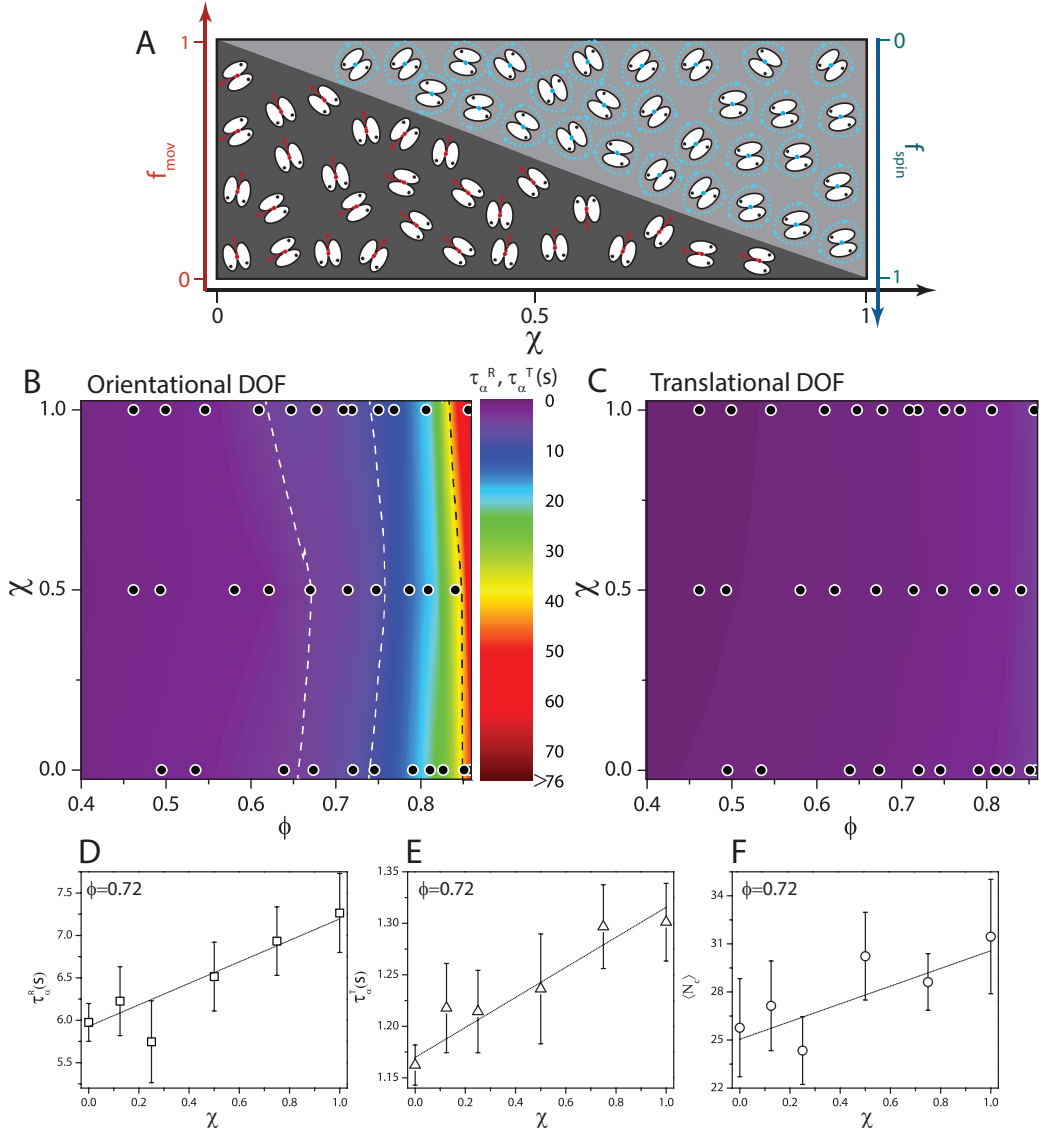}
\caption{\textbf {Relaxation dynamics of polar chiral active liquids depends on their net chirality, $\chi$.} \textbf{(A)} Schematic showing the maximum fraction of particles that can form a mover bond $f_{mov}$ and a spinner bond $f_{spin}$, at any given instant, versus $\chi$ for liquids of $\text{C}_2$ ellipsoids. \textbf{(B and C)} Relaxation dynamics phase diagram in ($\chi$, $\phi$) plane for orientational and translational degrees of freedom (DOF), respectively. The circles represent the $\chi$ and $\phi$ at which experiments were performed. In \textbf{B} the isochrones at intermediate $\phi$ are shown by white dashed lines and at high $\phi$ by a black dashed line. The color bar indicates the value of  $\tau_{\alpha}^R$ and $\tau_{\alpha}^T$. $\tau_{\alpha}$ for $\phi$'s in between experimental data points were obtained from linear interpolation. \textbf{(D and E)} $\tau_{\alpha}^R$ vs $\chi$ and $\tau_{\alpha}^T$ vs $\chi$, respectively, at $\phi= 0.72$ \textbf{(F)} $\langle N_{c}\rangle$ vs $\chi$ for $\phi$= 0.72. These clusters are of the top 20\% least-mobile particles over a time interval of 5$t^*$, where $t^*$ is the cage breaking time. In  \textbf{(D)} to \textbf{(F)}, the error bars represent standard deviation of the mean and were obtained from multiple statistically independent realizations of an experiment.}
\label{Figure3}
\end{figure}

Since the net chirality, $\chi$, of our active liquids constrains the fraction of particles that can form movers, $f_{mov}$, at a given time, even naively we expect that at fixed $\phi$, varying $\chi$ alone will alter their relaxation dynamics. In a racemic mixture, where $\chi = 0$, since every $(+)$ monomer can in principle be paired with a $(-)$ monomer, the maximum value of $f_{mov}=1$, while in an enantiopure liquid, where $\chi =1$, $f_{mov}\equiv0$. In fact, in liquids of $\text{C}_1$ and $\text{C}_2$ ellipsoids, where both movers and spinners exist, the consequences of changing $\chi$ on structural relaxation will only be amplified. Here, a decrease in $f_{mov}$ with increasing enantiomeric purity, is accompanied by a concomitant increase in the fraction of particles in the liquid that can make spinners, $f_{spin}$ (Fig. \ref{Figure3}A).

To begin with, we looked for quantifiable differences in the relaxation dynamics of liquids of  $\text{C}_2$ ellipsoids across three different net chiralities: $\chi = 1$ ($100\%$ $(+)$ monomers), $\chi = 0.5$ ($75\%$ $(+)$ monomers and $25\%$ $(-)$ monomers) and $\chi = 0$ ($50\%$ $(+)$ monomers and $50\%$ $(-)$ monomers). We varied $\phi$ from 0.1-0.84 while ensuring that across $\chi$, the $\phi$s were nearly identical to enable comparison. Enantiopure ($\chi = 1$) and enantiomeric excess ($\chi = 0.5$) liquids had to be made by manually placing monomers of the correct handedness, almost $8000$ particles for $\phi=0.84$, and this limited the number of ($\chi, \phi$) values we could study. For all $\chi$, we observed the formation of dynamic aggregates, reminiscent of motility-induced clusters, for $\phi\gtrapprox 0.2$ \cite{cates2015motility} (Fig. S9). There was no striking difference in the static or dynamical properties of these clusters with $\chi$ and we do not dwell on it here. Next, we quantified the structural relaxation time for the orientational degrees of freedom (DOF), $\tau_\alpha^R$, and the translational DOF, $\tau_\alpha^T$, through their respective time-correlators (Fig. S10). Like in passive liquids of elongated particles \cite{zheng2011glass}, even for our chiral active liquids, at fixed $\chi$ and with $\phi$, orientational slowing down is more rapid than the translational one - $\tau_\alpha^R$ grows by more than two orders of magnitude while $\tau_\alpha^T$ by one (Fig. \ref{Figure3}B, Fig. \ref{Figure3}C and Fig. S11). Remarkably, the isochrones (dashed white lines in Fig. \ref{Figure3}B) are not parallel to the ordinate but appear curved indicating that at constant $\phi$, changing $\chi$, alters orientational dynamics. In fact, signatures of a $\chi$-dependent relaxation persist even at $\phi\approx0.84$ (dashed black line in Fig. \ref{Figure3}B). Over the limited $\chi$ values studied, the same, however, cannot be said about translational dynamics.

To strengthen these findings, for $\phi = 0.72$, we carried out experiments for six values of $\chi$. The effect of changing $\chi$ on relaxation dynamics is now clear - both $\tau_\alpha^R$ (Fig. \ref{Figure3}D) and  $\tau_\alpha^T$ (Fig. \ref{Figure3}E) increase almost linearly with $\chi$, albeit the change in the latter is smaller than in the former. While it is tempting to reconcile these findings by a trivial decomposition of the liquids' relaxation into those arising from movers and spinners, we note that the relaxation of density fluctuations in dense liquids is often spatiotemporally heterogeneous and also collective. We borrowed from studies on supercooled liquids to quantify these dynamical heterogeneities in our active liquids and estimate their size \cite{weeks2000three,gokhale2016deconstructing}(see Fig. S12). At $\phi = 0.72$ and for all $\chi$, we indeed found density relaxation was dynamically heterogeneous and collective - the top $20\%$ of the least-mobile particles over a time interval $5t^*$ are spatially clustered. Here $t^*$, also called the cage-rearrangement time. For each $\chi$, we then determined the average cluster size $\langle N_c \rangle$, from the probability distribution of the cluster sizes. Remarkably, $\langle N_c\rangle$, which sets the time over which structure relaxes, also grows nearly linearly with $\chi$ for $\phi = 0.72$ (Fig. \ref{Figure3}F). 

\begin{figure}[htbp]
\centering
\includegraphics[width=0.99\textwidth]{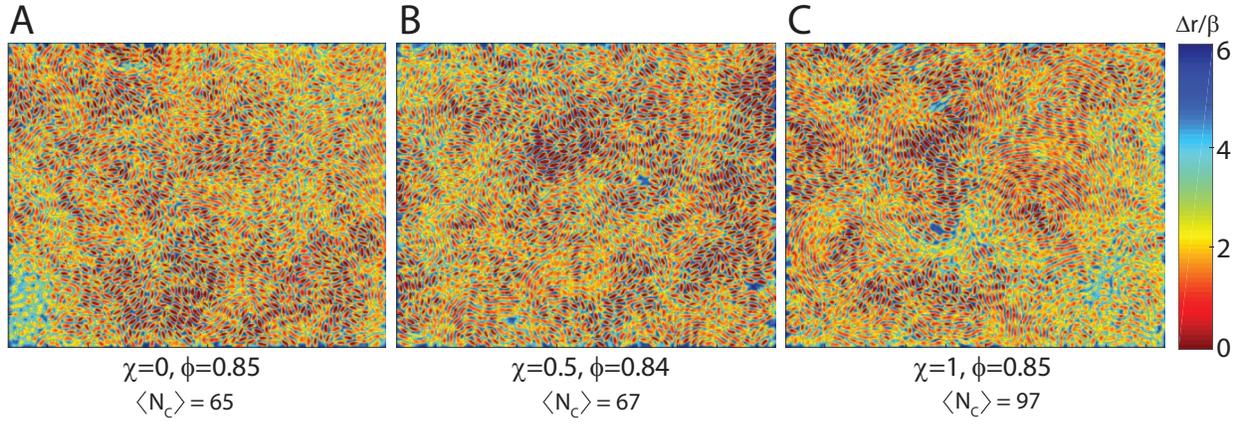}
\caption{\textbf {Chirality-dependent cooperative dynamics} \textbf{(A-C)} Particle displacement maps over  7$t^*$ for $\phi = 0.85$. \textbf{(A)} $\chi=0$ (racemic mixture). \textbf{(B)} $\chi=0.5$ (enantiomeric excess). \textbf{(C)} $\chi=1$ (enantiopure). Color bar shows the magnitude of scaled particle displacement $\Delta r/\beta$ over 7$t^*$}
\label{Figure4}
\end{figure}

The evolution in not just the size but also the nature of these dynamical heterogeneities with $\chi$ was even more striking for $\phi \approx 0.85$ 
(Fig.\ref{Figure4} and Movie S7). In the enantiopure liquid (Fig. \ref{Figure4}C), these heterogeneities were mostly large  spinning vortices comprising of almost $200$ particles, with the chirality of these vortices being the same as the monomers'. At $\chi =0$ (Fig. \ref{Figure4}A) and $\chi = 0.5$ (Fig. \ref{Figure4}B), we observed both ($+$) and ($-$) vortices and these were also smaller in comparison to $\chi = 1$. Besides vortices, we also observed streaming flows like those seen in liquids of achiral active particles \cite{thampi2014vorticity}. While orientational relaxation is hindered at this $\phi$, these vortices and streaming flows, which have no counterparts in equilibrium liquids, provide new pathways  \cite{berthier2014nonequilibrium} for translational relaxation of structure and is consistent with the observed weak growth in $\tau_\alpha^T$.  
 
Summarizing, our experiments on shape achiral grains with tunable chiral activity have helped uncover the purely dynamical analogues of stereoselectivity and chiral self-recognition. The dynamics of our chiral active liquids is exceptionally rich. Over a window of chiral activities, we observed the formation of homo- and heterochiral dimers with distinct dynamics and this influenced behaviour even in the dense regime. Most remarkably, varying only the total chirality of the active liquids from racemic to enantiomerically pure resulted in dynamical slowing down due to stark changes in the nature of collective relaxation. We anticipate that bringing in complexity in the particle through shape and/or internal degrees of freedom will have a profound influence on the emergent behaviour of chiral active liquids. Importantly, we have shown that even in systems, where the constituent particles lack attractive interactions \cite{lim2019cluster}, introducing chiral activity can bring about specificity in interactions and this can prove to be a powerful route for steering their self-assembly.

\bibliographystyle{apsrev4-1}

\begin{thebibliography}{10}
\bibitem{morrow2017transmission}
S.~M. Morrow, A.~J. Bissette, S.~P. Fletcher, Transmission of chirality through space and across length scales. {\it Nature nanotechnology\/}
  {\bf 12}, 410 (2017).

\bibitem{jedrzejewska2017making}
H.~Jedrzejewska, A.~Szumna, Making a right or left choice: chiral self-sorting as a tool for the formation of discrete complex structures. {\it Chemical reviews\/} {\bf 117}, 4863-4899 (2017).

\bibitem{kokot2017active}
G.~Kokot, {\it et~al.\/}, Active turbulence in a gas of self-assembled spinners. {\it Proceedings of the National Academy of
  Sciences\/} {\bf 114}, 12870-12875 (2017).

\bibitem{soni2019odd}
V.~Soni, {\it et~al.\/}, The odd free surface flows of a colloidal chiral fluid. {\it Nature Physics\/} {\bf 15}, 1188-1194 (2019).

\bibitem{grzybowski2002dynamic}
B.~A. Grzybowski, G.~M. Whitesides, Dynamic aggregation of chiral spinners. {\it Science\/} {\bf 296}, 718-721 (2002).

\bibitem{petroff2015fast}
A.~P. Petroff, X.-L. Wu, A.~Libchaber, Fast-moving bacteria self-organize into active two-dimensional crystals of rotating cells. {\it Physical review letters\/} {\bf
  114}, 158102 (2015).

\bibitem{jennings1901significance}
H.~S. Jennings, On the significance of the spiral swimming of organisms. {\it The American Naturalist\/} {\bf 35}, 369-378 (1901).

\bibitem{su2012high}
T.-W. Su, L.~Xue, A.~Ozcan, High-throughput lensfree 3D tracking of human sperms reveals rare statistics of helical trajectories. {\it Proceedings of the National Academy of
  Sciences\/} {\bf 109}, 16018-16022 (2012).

\bibitem{tsai2005chiral}
J.-C. Tsai, F.~Ye, J.~Rodriguez, J.~P. Gollub, T.~Lubensky, A chiral granular gas. {\it Physical
  review letters\/} {\bf 94}, 214301 (2005).

\bibitem{kummel2013circular}
F.~K{\"u}mmel, {\it et~al.\/}, Circular motion of asymmetric self-propelling particles. {\it Physical review letters\/} {\bf 110},
  198302 (2013).

\bibitem{workamp2018symmetry}
M.~Workamp, G.~Ramirez, K.~E. Daniels, J.~A. Dijksman, Symmetry-reversals in chiral active matter. {\it Soft matter\/} {\bf
  14}, 5572-5580 (2018).

\bibitem{scholz2018rotating}
C.~Scholz, M.~Engel, T.~P{\"o}schel, Rotating robots move collectively and self-organize. {\it Nature communications\/} {\bf 9}, 1-8
  (2018).

\bibitem{marchetti2013hydrodynamics}
M.~C. Marchetti, {\it et~al.\/}, Hydrodynamics of soft active matter. {\it Reviews of Modern Physics\/} {\bf 85}, 1143 (2013).

\bibitem{lowen2016chirality}
H.~L{\"o}wen, Chirality in microswimmer motion: From circle swimmers to active turbulence. {\it The European Physical Journal Special Topics\/} {\bf 225}, 2319-2331 (2016).

\bibitem{liebchen2017collective}
B.~Liebchen, D.~Levis, Collective behavior of chiral active matter: pattern formation and enhanced flocking. {\it Physical review letters\/} {\bf 119}, 058002 (2017).

\bibitem{lei2019nonequilibrium}
Q.-L. Lei, M.~P. Ciamarra, R.~Ni, Nonequilibrium strongly hyperuniform fluids of circle active particles with large local density fluctuations. {\it Science advances\/} {\bf 5}, 7423 (2019).

\bibitem{palacci2013living}
J.~Palacci, S.~Sacanna, A.~P. Steinberg, D.~J. Pine, P.~M. Chaikin, Living crystals of light-activated colloidal surfers. {\it
  Science\/} {\bf 339}, 936-940 (2013).

\bibitem{cates2015motility}
M.~E. Cates, J.~Tailleur, Motility-induced phase separation. {\it Annu. Rev. Condens. Matter Phys.\/} {\bf 6}, 219-244 (2015).

\bibitem{kudrolli2008swarming}
A.~Kudrolli, G.~Lumay, D.~Volfson, L.~S. Tsimring, Swarming and swirling in self-propelled polar granular rods. {\it Physical review
  letters\/} {\bf 100}, 058001 (2008).

\bibitem{deseigne2010collective}
J.~Deseigne, O.~Dauchot, H.~Chat{\'e}, Collective motion of vibrated polar disks. {\it Physical review letters\/} {\bf
  105}, 098001 (2010).

\bibitem{kumar2014flocking}
N.~Kumar, H.~Soni, S.~Ramaswamy, A.~Sood, Flocking at a distance in active granular matter. {\it Nature communications\/} {\bf
  5}, 1-9 (2014).

\bibitem{harris2015generating}
D.~M. Harris, J.~W. Bush, Generating uniaxial vibration with an electrodynamic shaker and external air bearing. {\it Journal of Sound and Vibration\/} {\bf 334}, 255-269
  (2015).

\bibitem{bechinger2016active}
C.~Bechinger, {\it et~al.\/}, Active particles in complex and crowded environments. {\it Reviews of Modern Physics\/} {\bf 88}, 045006 (2016).

\bibitem{fily2012cooperative}
Y.~Fily, A.~Baskaran, M.~C. Marchetti, Cooperative self-propulsion of active and passive rotors. {\it Soft Matter\/} {\bf 8}, 3002-3009
  (2012).

\bibitem{zheng2011glass}
Z.~Zheng, F.~Wang, Y.~Han, {\it et~al.\/}, Glass transitions in quasi-two-dimensional suspensions of colloidal ellipsoids. {\it Physical review letters\/} {\bf 107}, 065702 (2011).

\bibitem{weeks2000three}
E.~R. Weeks, J.~C. Crocker, A.~C. Levitt, A.~Schofield, D.~A. Weitz, Three-dimensional direct imaging of structural relaxation near the colloidal glass transition {\it Science\/} {\bf 287}, 627-631 (2000).

\bibitem{gokhale2016deconstructing}
S.~Gokhale, A.~Sood, R.~Ganapathy, Deconstructing the glass transition through critical experiments on colloids. {\it Advances in Physics\/} {\bf 65}, 363-452 (2016).

\bibitem{thampi2014vorticity}
S.~P. Thampi, R.~Golestanian, J.~M. Yeomans, Vorticity, defects and correlations in active turbulence. {\it Philosophical Transactions of
  the Royal Society A: Mathematical, Physical and Engineering Sciences\/} {\bf 372}, 20130366 (2014).

\bibitem{berthier2014nonequilibrium}
L.~Berthier, Nonequilibrium glassy dynamics of self-propelled hard disks. {\it Physical review letters\/} {\bf 112}, 220602 (2014).

\bibitem{lim2019cluster}
M.~X. Lim, A.~Souslov, V.~Vitelli, H.~M. Jaeger, Cluster formation by acoustic forces and active fluctuations in levitated granular matter. {\it Nature Physics\/} {\bf 15}, 460-464 (2019).

\end{thebibliography}

\section*{Author Contributions}
PA and RG conceived research and designed experiments. PA performed experiments and carried out data analysis. AKS contributed to project development and provided inputs with the manuscript. RG steered research and wrote the paper with PA.

\section*{Acknowledgments}
We thank Shreyas Gokhale for critical feedback on our manuscript. \textbf{Funding: }AKS thanks Department of Science and Technology (DST), Govt. of India for a Year of Science Fellowship. RG thanks DST-SwarnaJayanthi fellowship (2016-2021) for financial support.
\textbf{Competing interests:} The authors declare that they have no competing interests. \textbf{Data and materials availability:} All data needed to evaluate the conclusions in the paper are present in the paper and/or the Supplementary Materials. Additional data available from authors upon request.

\newpage
\section*{Figures and Tables}

\begin{table}[htb]
\centering
\begin{tabular}{cccccccc}
\noalign{\smallskip} \hline \hline \noalign{\smallskip}
$ \text{Label}      $ & $\alpha$  &   $\beta$    & $\gamma$      & $\Delta m_{LR}$ & $\omega$     & $R$     &     Dimer state observed\\
$      $ &         mm          &         mm              &            mm  &          g            &          rad/s              &           mm \\
\hline
$\text{C}_{1}$      & 6      & 3   &  2.1  & 6.91 x $10^{-3}$    & 12.2   &  1.3   &  Spinners, Movers \\
$\text{C}_{2}$      & 6      & 3   &  2.1  & 5.85 x $10^{-3}$    & 11.9   &  1.4   &  Spinners, Movers \\
$\text{C}_{3}$      & 6      & 3   &  2.1  & 4.78 x $10^{-3}$    & 8.5    &  2.2   &   Movers \\
$\text{C}_{4}$      & 6      & 3   &  2.1  & 3.13 x $10^{-3}$    & 5.1    &  4.1   &  Movers \\
$\text{C}_{5}$      & 6      & 3   &  2.1  & 1.97 x $10^{-3}$    & 2.5    &  6.1   &  Movers \\
$\text{C}_{6}$      & 6      & 3   &  2.1  & 1.04 x $10^{-3}$    & 2.1    &  10    &  Movers \\

\noalign{\smallskip} \hline \noalign{\smallskip}
\end{tabular}
\caption{Details of various 3D-printed polar chiral active ellipsoids. }
\label{Table1}
\end{table}

\end{document}